\documentclass{article}
\input{epsf}

\def\ct{\cite}
\def\be{\begin{equation}}
\def\ee{\end{equation}}
\def\bi{\bibitem}

\begin{document}
\date{August 11, 2000}

\title{\bf the quantum vacuum, fractal geometry, and the quest for a new theory of gravity}
\author{Edward F. Halerewicz, Jr.\footnote{email:  ehj@warpnet.net} \\ \\ Lincoln Land Community College\footnote{This address is given for mailing purposes only, since I'm  a student and do not hold a professional position at the above address.  This work was made possible through my own personal research and studies, and does not reflect the above listed institution.}
\\ 5250 Shepherd Road\\ Springfield, IL 62794-9256 USA\\}
\maketitle

\begin{abstract}
In this letter recent developments are shown in experimental and
theoretical physics which brings into question the validity of
General Relativity.  This letter emphasizes the construction of a
fractal $3+\phi^3$ spacetime, in $N$-dimensions in order to
formalize a physical and consistent theory of `quantum gravity.'
It is then shown that a `quantum gravity' effect could arise by
means of the Strong Equivalence Principle.  Which is made possible
through a pressure of the form $-\kappa (R^{ca}_{b}-{1\over
2}g^{c\sigma}_{ab}R^c)=\kappa T^{c\sigma}_{ab}$.  Where it is seen
that nuclear pressures can be added to the gravitational field
equations by means of twistor spaces.
\end{abstract}



\section{Introduction}
Recently the validity of General Relativity (GR) has been brought
to question by Yilmaz \ct{1}, et al.  Although such
interpretations allow for gravitation to be mathematically
consistent and singularity free.  Such revisions fail to describe
the behavior of test particles as adequately as GR, elevating GR
as the correct theory \ct{2}.  Today certain questions about GR
remain relevant, such as how does it relate to vacuum energy and
quantum mechanics in general.  It has been shown in previous works
that GR remains self consistent when including the quantum vacuum,
or zero-point field \ct{3}.  However, the search for a self
consistent theory of ``quantum gravity," remains a major
theoretical challenge today.  Among the theoretical arguments
against the standard interpretation of GR is the choice of
mathematical coordinate systems \ct{4}.  Special Relativity (SR),
is based upon the structure of a flat Minkowski spacetime given in
a four-dimensional coordinate system.  Recently attempts have been
made describing coordinate systems with fractal spaces as opposed
to natural ones \ct{5, 6, 7}.  Such an adaptation as the case with
the Yilmaz approach eliminates singularities within the field
equations.

Recent observational and experimental data have also put into
question  the validity of GR.  The National Aeronautics and Space
Administration (NASA) has reported an anomalous acceleration of
$\pm 8.5 \times 10^{-8}cm$, on spacecraft on the outer edge of the
solar system \ct{8}.  This data was obtained from information
gathered by the Jet Propulsion Laboratories (JPL), and the Deep
Space Network (DSN).  Thus far, no satisfactory conclusion has
been given to explain the so called ``anomalous acceleration
towards the sun."  Not only have spacecraft provided some
fundamental flaws with gravitation, but laboratory results as
well.  Dr. Eugene Podkletnov, has reported a ``gravitational
shielding" effect with composite bulk $YBa_2 Cu_3 O_{7-x}$ ceramic
plates \ct{9}.  In light of all of these developments it is hard
to consider GR as the correct theory.  It is the opinion of the
author that GR is a theory that ``works," however it doesn't
necessarily make it the correct theory.

The goal of this letter is to show that GR is not the correct
theory of gravitation, but just works exceptionally well.  Just as
previously Newton's law of universal gravitation worked
exceptionally well.  This letter is not intended to be a
replacement for GR, nor is it intended to present theoretical
flaws of the that theory.  This letter is only presented as an
introductory work for an alternative theory of gravitation.  The
general theme of this letter is given by the following postulates:

\newtheorem{post}{Postulate}

\begin{post}
(virtual gravitation).  Spacetime is not a
null energy field, it consist of asymptotic vacuum fluctuations,
and behaves as a virtual ``energy-sheet."
\end{post}

\begin{post} (planckian invariance).  The Planck length
is a gauge invariant function for all (interacting) brane
observers. [an adaptation to the postulate of New Relativity.]
\end{post}

This letter is presented in the following format in section
\ref{uni} a brief introduction into unified field theories are
given.  In section \ref{qgh} a few quantum gravity approaches are
introduced.  In section \ref{fg} fractal geometry is introduced
and its relations to a complex system are given.  In section
\ref{qedf} the meaning of fractal geometry for QED is discussed.
In section \ref{qcdf} the meaning of a fractal geometry is
discussed for QCD.  In section \ref{card} a new theoretical
particle is introduced utilizing fractal geometry.  In section
\ref{flt} a relationship between N-dimensional and two-dimensional
systems are given.  In section \ref{geo} a philosophy of geometry
is given.  In section \ref{vac} the effects of the quantum vacuum
are discussed.  In section \ref{stg} a relationship between
fractal geometries and the quantum vacuum are discussed. In
section \ref{feystg} the meaning of Feynman diagrams are
discussed.  In section \ref{qm} the validity of quantum mechanics
is brought into question.  In section \ref{bm} an alternative
description of gravity is given which may explain the EPR paradox.
In section \ref{cqg} an overview of a canonical non Riemannian
gravitational field is given.  In section \ref{can} the planck
length results as a function in canonical quantum gravity.  In
section \ref{aa} a possible alternative for the ``anomalous
acceleration" of spacecraft is given.  In section \ref{pgeo}
pseudo geodesic equations are presented.  In section \ref{dis} a
general discussion of this work is presented.  In section \ref{pl}
a discussion of the meaning of the planck length is given.  The
conclusions of this work is drawn in section \ref{con}, which
gives stronger definitions to equivalence principle in Appendix
\ref{ep}.  Finally it is suggested that there may exist a
detectable from of ``Yang-Mills Gravity" in Appendix \ref{ymg}.

\section{Unification a brief history}\label{uni}

\begin{quote}
``{\it I am convinced that He [God] does not play dice.}" \flushright--A. Einstein
\end{quote}

\begin{quote}
``{\it Einstein, quit telling God what to do.}" \flushright --N. Bohr
\end{quote}

The unification of gravitation with quantum mechanics began with
Einstein's objections to the newly developed quantum theory.
Although acknowledging the successes of the new theory he believed
it to be incomplete.  Einstein was convinced that there was a
deeper theory involved, one which would also include GR, a unified
field theory was christened.  Soon came the work of Kaluza and
Klien, giving a pseudo mathematical unification of
electromagnetism and gravitation.  The theory would soon die out
and loose interest, until quantum mechanics came around.  And
asked the simple question, how does gravity behave at the quantum
level; the answer Kalzua-Klien gravity \ct{10}.  Research in this
area soon exploded, extra dimensions were soon added to the field
equations, superstring theory was born.  Particle physics began
unifying fundamental forces as well, the weak force, the strong
force, the electromagnetic force.  But no gravitational force,
Cosmologists helped out, the big bang and nucelarsynthesis would
help to explain the problem.  Soon physics became littered with
Grand Unified Theories (GUTs) and Theories of Everything (TOEs),
they all have the approach of  a ``unified field theory." However,
they missed the simple point Einstein was trying to make, how do
quantum mechanics and relativity relate?  This is a hierarchical
question, the relevant question is how do macroscopic and
microscopic worlds communicate?

\subsection{quantum gravity}\label{qgh}
Historically there have been two models formulated for the
construction of a consistent quantum gravity theory.  They are the
canonical and Hamiltonian approaches \ct{11}.  These two
approaches have had limited success, however more recently the
theory of loop quantum gravity \ct{12} has been introduced into
the sea.  Out of the three approaches presented loop quantum
gravity is generally accepted as the correct approach. However,
for a more accurate description of the historic developments of
quantum gravity see \ct{13}.  In this letter I will focus on the
canonical approach as it relates to gauge invariance.

\section{Fractal Geometry}\label{fg}
\begin{flushleft}
``{\dots\it If we're built from Spirals while living in a giant
Spiral, then is it possible that everything we put our hands to is
infused with the Spiral?}"  \\\flushright --Max Cohen in the motion picture $\pi$
\end{flushleft}

The presence of matter within GR disturbs the field equations by
the existence of singularities or ``point-particles."  How can one
avoid this eye sore in the equations, simple fractal geometry.  If
matter is fractal it can not condense into points, however this
allusion can still take place above the planck energy scale.  It
is hard to believe that this simple approach has only been
attempted in recent times, fractal sets are more common in nature
than simple polygons. First let us begin with a simple
construction of a fractal set with the simple equation $Z_{(n)} =
(Z_{(n - 1)} )^2  + C$.  One must also realize that a fractal is
composed of a complex number system, i.e. a+ib.  Using this form
one may wish to construct a complex averaging of the mean, which
results from the golden mean $\phi=\sqrt {5-1}/\sqrt {(5-1)}$.
Thus we have constructed a complex mean which has two possible
solutions as seen below:
\be
C_m  = \frac{{\sqrt {a - 1} }}{{\sqrt {(a - 1)} }} + \frac{{\sqrt { - b - 1} }}{{\sqrt {(b - 1)} }} = \left\{ \begin{array}{l}
 {{{\sqrt {a + 2} } \over {\sqrt {(a + 2)} }}};{{{\sqrt { - b} } \over {\sqrt {(b)} }}} \Leftrightarrow a = c \\
 {{{\sqrt a } \over {\sqrt {(a)} }}};{{{\sqrt { - b + 2} } \over {\sqrt {(b + 2)} }}} \Leftrightarrow a \ne c \\
 \end{array} \right.
\ee this complex mean thus has non communcating solutions.  Which
from the stand point of imaginary numbers yields the general
statements:
\be
\frac{{\sqrt { - n} }}{{\sqrt {(n)} }} = i \leftrightarrow
\frac{{\sqrt {(n)} }}{\sqrt{ - n}} = - i \ee Thus these statements
would appear to be in agreement of the theory of quaternions.
Which is interesting enough in itself, a four-dimensional version
of the complex number system.

With this preliminary work set we can now construct a Minkowski
spacetime which that takes advantage of fractal dimensions.  First
one can construct the generalized three-dimensional manifold as a
three-brane, and thus incorporating time as a fractal set.  Thus
the fractal construction on spacetime is presented in the form
$3+\phi^3$ a similar approach was made in Ref. \ct{4}.  It is
here postulated the origin for this three-dimensional brane arises
from planck scaling.  The reason for this postulate is seen when
the golden mean is applied to $N$-dimensions $\phi
_{n}=\sqrt{N-1}/\sqrt{(N-1)}$, the higher n the closer the
mean is to 1.  Thus only when $n$ approaches infinity will we see
that $3+\phi^3$ will yield the standard Minkowski space of 3+1
dimensions.  This is of course in agreement with any system that
we apply c=$\hbar$=$\kappa$=$K$=1 more evidently this corresponds
to a time dilation effect in terms of special relativity.  Such
that we have the following revision to the flat four-dimensional
Minkowski space:
\be
(\omega, z^2)=\omega c^2-(\phi)z^2_{1}-(\phi)z^2_{2}-(\phi)z^2_{3}
\label{k}
\ee
It is interesting to note that this pseudo metric
appears to be an inverse of the standard Minkowski spacetime, this
importance is seen in section \ref{bm}.  So far this method has only
left intriguing consequences, however it diverges from the point
made earlier in this section.

In string theory the atomism view of matter is replaced with
vibrating strings, these vibrating strings correspond to a real
geometry.  However, fractal geometries are allowed to break of
these strings into imaginary components.  These imaginary
components thus make the string a complex function, yielding a
pseudo point-like string.  To analyze this premise allow us to
view the Nambu-Goto action $S=\mu _{0} \int d^2 \xi \sqrt{g}$.  If
strings can indeed be made to subside with fractal geometry, then
they would break off into imaginary components by the empirical
action:
\be
S=\aleph _{0} \int\omega ^2 \phi \sqrt{-g} \ee where
$g=\lambda{^2} dz^2\otimes d \bar{z}^2 \equiv x+iy$.  This pseudo
string forms many more string components in $N$-dimensions, that is
the string fragment continues in infintium.  However, in the
physical sense the string exist a pseudo point-like particle, do
to the scaling nature of the planck length.  These fractal strings
then interact within a field, known as gravitational or zero-point
fields.  When the fractal strings converge with other fractals, a
self organization takes place, i.e. the production of virtual
particles.  This production is made possible through the non
communacating mathematics associated with quantum mechanics. As
the particles are produced they destroy one another, such that
their world-sheets reverberate in a complex form.  This complex
reverberations in $N$-dimensions is responsible for the vibration of
the string, which we na\"\i vely associate with mass.

\subsection{QED fractality}\label{qedf}
Electromagnetic waves are the result of four-dimensional
interactions, and its real wave would correspond to the results
found in Classical Mechanics.  However, it would intern have a
fractal complex field, which would cause the field to break
periodically yielding a string fragment, a quanta.  This quanta in
non self regulating, i.e. it is the nature of the real wave which
causes the string to reverberate.  The above consideration may
carry some controversy in the well known theory of Quantum
Electrodynamics (QED).  If a quanta is just a fractal string, then
what is the proper approach for the exchange of energy between the
two systems?  Well, the result appears as classical approximation,
the quanta hits the string as a solid body, causing a change in
geometry, which virtual particles oppose.  This causes the string
to ``bounce" back to its original form, emitting a real wave, but
not necessarily a quanta, remember a quanta is  given by a complex
field.  Since quantum mechanics is sketched out onto a point
particle-like environment, its consequences would agree with the
QED model.  In fact the fractal model yields a much physical
picture for non communacating relationships than the quantum
theory.

\subsection{QCD fractality}\label{qcdf}
The above result would agree with QED, however, its definitions
are quite weak, in fact one may expand these definitions to
Quantum Chromodynamics (QCD).  Thus a nucleon may be made to
reflect electromagnetic radiation as well, however, this dose not
defy the documented experiments in any magnitude.  It would be the
interaction of the system, i.e. what string perimeters give way to
the colored, and other gauged forces, that yields its
``particles".  Each string fragment consist of its own local
vibrations (gauge invariance), which attributes its mass, i.e.
differentiates between a Higgs particle and a quark.  These states
then have their own local statistics, their real waves would then
correspond differently than the electromagnetic field.  Which
results in the production of the celebrated Yang-Mills field, and
thus yields the production of colored particles such as the gluon.

\section{Cardinal Strings}\label{card}
Cantor pioneered the study of infinities with his new theory of
Cardinal numbers, however he faced opposition in his time for this
new theory \ct{14}.  Cardinal numbers offer the best insight into to
the study of fractal strings in N-dimensions, these
interpretations in fact have direct physical consequences.  Most
notably it can explain the situation unleashed by the infamous
EPR-Bell paradox, where Faster Than Light (FTL) communication
appears possible (depending on planckian scaling).  For recent
theoretical implications and interpretation of the EPR-Bell
paradox see \ct{15}.  In each scaling the laws of physics would be
very different, and hence superluminal velocities would seem to
appear in lower branes.  Thus the traditional interaction of
strings should not be limited on a specified dimension, but behave
as a set of Cardinal numbers.  Thus a slight revision of the
Nambu-Goto action should be given which yields
\be
S=\aleph _{0} \int\omega ^2 \phi \sqrt{-\tilde g}
\ee
where \(\tilde g=-dt' \otimes dt'+[dr' \otimes dr'+{1 \over 4}(sin(2r'))^2 \Omega -{2}]\).

\subsection{Fermat's Last Theorem}\label{flt}
\begin{flushleft}``{\it To divide a cube into two or other cubes, a fourth power, or, in general, any power whatever into two powers of the same denomination above the second is impossible}\dots" \\ \flushright --Fermat
\end{flushleft}

Fermat's Last Theorem can be associated with fractal geometry in
one respect, there is no general real solution to fractal geometry
above dimension 2.  This may be a simple coincidence and may have
no deeper meaning, however, this is contrary to the recently
created Holographic Principle (HP) \ct{16}.  The HP relates that
the Universe may exist in dimensions of infinitum status. However,
the laws of physics are best projected onto a pseudo
two-dimensional screen, and our three-dimensional world is only a
pseudo manifestation of an $N$-dimensional continuum.  In string
theory we can view our universe as made up of two two-dimensional
branes (described by type IIA D2 membranes).  Thus any other
dimension outside the holographic conjecture yields no physical
meaning and no solution.  Just as what is suggested by Fermat's
Last Theorem, therefore our four dimensional slice of the brane is
a pseudo physical manifestation of the holographic screen.  There
is only one explanation for this result, there must exist a
physical constant for specified energy scales, i.e. the planck
scale.  Here another coincidence appears to arrive, the two
dimensional wave equation for string theory $\left( {{{\partial
^2 } \over {\partial \sigma ^2 }} - {{\partial ^2 } \over
{\partial \tau ^2 }}} \right)\chi ^\mu  (\sigma ,\tau ) = 0$.  It
seems that both mathematically and physically there is a special
importance with dimension 2.  This discussion is largely
philosophical, however it is interesting to note that \it cardinal
strings \rm are given by complex numbers.  In fact a cardinal
string in four-dimensions is remarkably similar to the
two-dimensional form, and seems to correspond to a torus:
\be
S^2=\aleph _{0} \int d^2 \xi\sqrt{-g}\,+\aleph _{0} \int d^2
\xi\sqrt{-g}\equiv \aleph _{0} \int d^4 (x{1 \over
\lambda^2_{IIA}t^2_{IIA}}\cal M) \ee From this complex
structuring, and properties of cardinal numbers it can be seen why
quaternions were alluded to in section \ref{fg}.

\section{a matter of geometry}\label{geo}
Einstein's theory of GR transformed Newton's theory of a
gravitational force, to a direct consequence of geometry. However,
although the idea of a force was replaced with a geometry, the
geometry still yields a force when explained in Riemannian
geometry.  An even more radical approach to gravitation as a
geometry was produced by Roger Penrose in his theory of twistor
spaces.  The geometry itself is more important than physical
masses, in fact masses only come important when one expands this
theory.  This is true for a fractal revision of string theory, it
exists as  a pure geometry, the interacting geometry in fact produces
mass.  This seems almost a radical stance from the point of view
of GR, however geometry remains a key factor as the ideal of a
Force to GR.

\subsection{the vacuum}\label{vac}
The vacuum exist from a state of virtual particles being produced
via cardinal string fragments.  Since virtual particles are a pure
construction of ``particles" in $N$-dimensions, they are not true
strings (i.e. they violate the HP).  These particles thus carry no
mass-energy equivalent within our universe, never the less they still
posses a geometry.  This situation only holds true when the system
is localized, however, when interacting with non cardinal strings
can induce an energy exchange.  By the well known Casimir effect
the energy of the vacuum should be given by \ct{20}: $$ \rho _{ZP}
(\omega)={\hbar \omega ^{3} \over 2 \pi ^{2}c^{3}} $$ and when
interacting with an inertial mass system we have: $$ m_i={V _0
\over c^2 } \int \eta (\omega)\rho _{ZP}(\omega)d \omega. $$  This
relates the fact that as mass is accelerated it pushes the quantum
vacuum energy (which is analogous to the assumption made be
postulate one).  In other words it reacts in the same fashion as
air molecules do when inertial masses accelerate on earth
(producing pressure on the system).  Furthermore, it can be
assumed that a material body increases its rest mass by absorbing
this zero-point-energy (this assumption must be given in order to
satisfy conservation laws).

Moreover, since they are cardinal strings they are unified in a
manner, thus the vacuum is a geometrical manifestation of string
particles.  That is the geometrical patterns formed through string
interactions is what we call a gravitational field, i.e. a virtual
gravitational field.  Since these string interactions are only
virtual there is no reason to modify the Einstein Field Equation,
unless one wishes to discuss quantum string effects.   Furthermore
$N$-dimensional spacetime metrics have shown to be very similar to
the structure of four-dimensional spacetimes \ct{17}.  Therefore
the classical gravitational field is removed from quantum
mechanics as it exist in a virtual sense, thus quantum mechanics
is a property of matter, i.e. interacting geometries.

\subsection{the stage}\label{stg}
The vacuum however is not currently treated as the geometry of a
fractal spacetime system, and hence is incompatible with other
vacuum theories \ct{18, 19}.  However a fractal model for quantum
mechanics appears to agree with at least one interpretation of the
quantum vacuum \ct{20}.  In fact this interpretation goes right
along with loop quantum gravity, and string theory see Ref.
\cite{21}.  The Hausdorff (or fractal) dimension suggest that
dimensions maybe confined to a $D=3$ spacetime, with fractal string
scaling.  This principle was postulated earlier in this letter as
the consequence of the ``planckian scaling," and is the leading
postulate in New Relativity.  Here the importance now becomes what
is the meaning of de-Broglie phenomenon.  The Einstein de-Broglie
equation $\hbar \omega _{\bf C}=m_0 c^2$ can be seen as a
representation of the relativistic wave equation, i.e. mass is a
quantifiable measure of energy.  Which can be applied directly to
string theory, the vibrations of the string are given in a fractal
frequency comparable to the Compton wavelength.

\subsection{the Feynman stage}\label{feystg}
>From the above consideration it can be equally applied the the
origin of a bodies mass intern comes from the gravitational field
itself.  This requires the use of Feynman diagrams, and believe it
or not this approach is indeed correct, if strings are represented
by cardinal numbers (and if quantum mechanics is considered to be
correct).  Since this allows for FTL communication at the
classical level it can be interpreted at least at the quantum
level that mass originates from spacetime (when measured at the
planckian scale).  In fact at the quantum level the production of
virtual particles may be responsible for a light paths geodesic
curvature, yielding a quantum gravity theory.  More correctly it
may be viewed that the quantum theory is in reality a classical
approximation of string theory.

\newtheorem{theorem}{Theorem}
\begin{theorem} (Brussels approximation). Quantum mechanics
exist as a classical approximation of string interactions which
possesses an apparent time reversed symmetry.
\end{theorem}

\subsection{quantum mechanics?}\label{qm}
Deriving this theorem let us consider the following thought
experiment:  If mass is composed of vibrating stings, and intern
these strings produce gravitational fields in N-dimensions then
space is vibrating.  However, such an approach would imply that
geometrically speaking the two systems are unaware of their own
vibrations under gauge invariance.  On the other hand, the
interaction of fractal strings are given in a complex field, which
itself is anti-communcating.  Thus spacetime, or string space is
subjected to Uncertainty Principles as shown in Ref. \ct{22}.
Since at the classical level, the fractal space can give way to
real solutions I now make the assertion that the quantum particles
are at flux, and not the space itself (for argumentative purposes
only).  Therefore when a quantum is observed, it is the space
which becomes ``fuzzy," not the particle, and when not observed
the inverse follows.  Thus the theorem leads to two possible out
comes during an observation sequence. i)  the fuzzy quantum
particle becomes a point particle, when time symmetries are
reversed.  ii) spacetime is fuzzy, however when collapsed by a
point particle elucidates to a ``natural" state.  Thus it is seen
that only when time symmetries are reversed does one obtain the
laws commonly associate with quantum mechanics.  Making the only
valid approximation of quantum mechanics the Brussels
Interpretation \ct{23}, this may also explain the EPR paradox.
However our thought experiment does yield one solution which
interpretation i and ii are consistent.

When a particle is observed by a frame, it is in reality observed
by the local states (brane) of the cardinal string, which may be
in any number of states.  As a point particle (non local string)
enters the system it begins to collapse the wave function of the
local state.  This makes it appear that a fuzzy quantum particle
has entered the system, much as a star appears to twinkle in the
night sky.  This collapsing of the state I will call the
observational's frame ``present," before the action the
observational state was fuzzy.  However, after the event a self
organization took place, an event occurred, producing a present
state.  After the quanta is observed by the observational frame,
its present state then becomes certain in the terminology of
classical quantum mechanics. Therefore local brane
string interactions can not take place until a non local
(cardinal) string collapses the wave function of the system.

\subsection{Bohmian Mechanics, gravitation, and EPR}\label{bm}
>From the Feynman interpretation of the time reversed symmetries of
the gravitational field, new conclusions about the nature of
spacetime can be made.  It is here postulated that classical
mechanics is in reality a description of a quantum system given
under an approximation of a time reversed symmetry.  Such that the
following statements become true:

\begin{itemize}\item Reversal of Bohmian Mechanics (BM) yields Classical Mechanics
\item Reversal of Brussels Interpretation (BI) yields Standard
Quantum Mechanics (SQM)
\end{itemize}

With the fractalization of spacetime given in section \ref{fg}, we may
conclude that (with the use of Bohmian Mechanics) that inertial
mass yields an expansion of spacetime.  Thus as a body gains mass
as it accelerates in classical spacetime it causes the
fractalization of the Bohmian system to increase (which is
analogous to a Lorentz transformation).  Which thus gives the
allusion that spacetime is contracting in the classical real
frame.  The gravitational force, thus is an inertial
acceleration which radiates a pseudo center of
gravity vector in terms of Newtonian mechanics.  However, this is
how we interpret the events classically, in reality it is the
expansion of the fractal Bohmian space (the $\phi^3$ term in
section \ref{fg}, e.g. the time dimension) which yields inertial
acceleration.

\begin{figure}
\centerline{\epsfbox{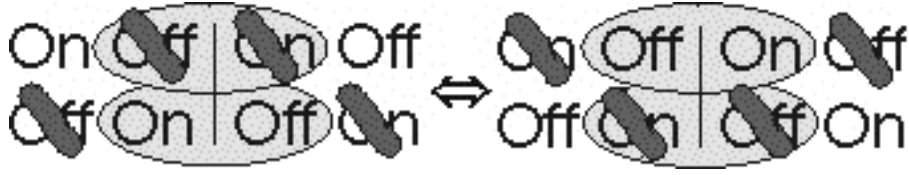}} \caption{An over
simplification of the EPR statistics for a spin ${1 \over 2}$
system given in a binary plane with time reversed symmetries.
Where the ovals represent the probability of being observed by the
apparatus.  And where the positions ``on" or ``off" represent the
outcome of events interpreted by classical mechanics.}\label{fig}

\end{figure}

Therefore, it is the Brussels interpretation of quantum mechanics
which yields the EPR paradoxes, the connection of the particles is
created by the (incorrect) approximation of time reversed
symmetries. That is to say the EPR paradox only includes simple
(non fractal) states, which by time reversal appears to yield FTL
communication (see figure \ref{fig}).

This appearance of FTL communication shouldn't be taken to
seriously since recent experiments (cfr. Wang, et al \ct{33})
appear to yield FTL communication.  However, it is the string
interactions which yield quantum mechanics, in fact it yields the
same interpretations as Bohmian Mechanics \ct{24}.  Thus Bohmian
Mechanics adequately describes the behavior of ``particles" while,
the Brussels Interpretations yields standard quantum mechanics
[meaning that this system is only an approximation].

\section{quantum gravity?}\label{cqg}
Since complex spaces have been presented as a solution to the
singularity problem, it is natural for a formulation of a complex
spacetime.  To proceed in this manner one must neglect the
cherished Einstein-Hilbert action $s=\int d^4 x\sqrt{-g}R$ and
replace it with the Tucker-Wang action: $$ s=\int \lambda ^2
R\star 1 $$ Therefore we can now discuss a complex gravitational
field, without the traditional Riemannian geometry.  The classical
Einsteinian relativity gives the generic field for a spacetime
geometry as $\nabla ^\mu G _{\mu \nu}=0$.  Such that I now wish
to make the generalized statement $\nabla ^{\hat\mu} G _{\hat\mu
\hat\nu}\ne 0$, or in canonical terms $\nabla _a G^{ab}=0$. As
such a generalization of a purely idealistic spacetime governed by
perfect fluid becomes:
\be
G^{ab}=16\pi GT_{ab}
\ee
where under ideal cases one can have the geometry $R^a_b +{1 \over 4}g_{ab}R$.  The reason the field
takes on the term $R^a_b$, as opposed to $R_{ab}$, can be seen
with the use Riemannian metrics.  First, let us begin with a Ricci
symmetric tensor of the form: $$ {{\partial \Lambda _\mu ^\sigma }
\over {\partial x^v }} = \eta _{\mu \nu \sigma } \Phi _v \Lambda
_\mu ^\sigma $$ which within a constant field becomes $R_{\mu
\nu}=\Lambda ^\sigma_\mu F_{\sigma \nu}$.  This field can thus
transpose to $F_{\mu \nu}=\partial _\mu  \Phi _\nu  -\partial
_\nu \Phi _\mu$, and couple to an opposing electromagnetic field
by the connection $$ \Gamma _{\mu \nu }^\sigma   = \Lambda _\mu
^\sigma  \Phi _\nu $$ Which therefore leads to the following
antisymmetric Riemannian field $$ R_{\mu \nu } \left( {{{{\partial
\Phi _v } \over {\partial x^\sigma  }}} - {{{\partial \Phi _\sigma
} \over {\partial x^\nu  }}} + C_{\nu \sigma }^{\alpha \beta }
\Phi _\alpha  \Phi _\beta  } \right)\Lambda _\mu ^\sigma   =
F_{\sigma \nu } \Lambda _\mu ^\sigma $$ Whence therefore means
that there must be an equivocal orthonormal action taking place,
such that one has $R_b^a  = d\Lambda _b^a +\Lambda _c^a \wedge
\Lambda _b^c$. In which the generic geometry for a perfect fluid
becomes that of
\be
G^{ab}=16 \pi GT^\sigma _{ab}
\ee
which translates to the field equation:
\be
R^a_b -{1 \over 4}g(\hat e_{(a)}) ,\hat e_{(b)})R=-{16 \pi G \over
c^4}T^\sigma _{ab} \ee This equation must be modified when given
in an $N$-dimensional system such that on has:
\be
R^a_b -{1 \over 2(n)}g(\hat e_{(a)},\hat e_{(b)})R=-{8 \pi(n)G
\over c^{2(n)}}T^\sigma _{ab} \label{wfe} \ee The above equation
is in essence a canonical gravitational field equation, which
appears to be a good candidate for a quantum gravity theory. Where
the geodesic equations become
\be
{d \over {d\hat l}}\left[ {\left( {1 + {{{\gamma _{ab(n)} } \over
{2(n)}}}} \right){{{dx^\mu  } \over {d\hat l}}}} \right] - \Gamma
_{[\not \beta \not \lambda ]}^{\not \alpha } {{{dx^{\not \beta } }
\over {d\hat l}}}{{{dx^{\not \gamma } } \over {d\hat l}}}\left( {1
+ {{{\gamma _{ab(n)} } \over {2(n)}}}} \right) \ne 0 \ee However,
this interpretation suggest that spacetime is quantitized by a
canonical action however, quantum particles are given as classical
particles.  Hence this interpretation would be an inverse of
understood quantum mechanics.  However, here a paradox opens up,
when time is reversed particles remain in one quantum state, thus
SQM is not retrieved.  Thus, it is seen that there exist no true
``quantum gravity" theory.  In fact if one applies this
formulation with the planck length it destroys the principle of
\it planckian invariance \rm and gives an allusion to the
existence of an \AE ther.

\subsection{canonical approach fails}\label{can}
Therefore the planck length no longer remains a constant but becomes a
dynamical function.  First let us write the planck length in terms
of $N$-dimensions and apply it to the above field equation such that
we have:
\be
l_{p \kappa}=(\kappa \hbar ^n /m^n _p c^{3n})^{1/2n}\cdot\psi.\ee
Momentum must be reevaluated from $E=\pm(p^n c^n+m^n _0
c^{2n})^{1/n}$, such that $m^n _{p_0}=\mp(p^n
c^n/c^{-2n})^{1/n}$.  Thereby the planck length, and mass are
actually given by a particles rest momentum.  Such that the planck
length is in reality given by the function:
\be
l_p =\pm ({\kappa \hbar ^n /\gamma m_{p_0 }^n
c^{3n}})^{1/2n}\cdot \psi. \ee This thereby has major
implications, that the planck length is not really a constant at
all but a function of momentum.  Such that as an object increase
in speed with respect to its rest momentum, its planck energy
becomes larger. That is as a material body is subjected to length
contraction, its planckian energy is modified to compensate for
the effect.   Since the momentum is measured at rest m remains a
constant, it is the velocity of the system which changes. Thereby
meaning that length contraction in special relativity is not given
by Lorentz transformations, but by the local rest momentum of the
planck barrier.  This results when we interpreted this action
canonically however under BM it yields expected results. Therefore
it is seen that a canonical formulation fails to keep ``planckian
invariance" which represents a failed attempt at a quantum gravity
theory.

\section{anomalous acclerations?}\label{aa}
Finally I bring light to an alternative explanation for the
acceleration of spacecraft \ct{8}.  Since the findings of the
``anomalous acceleration towards the sun," there have been a
number of possible explanations given \ct{27, 28, 29}.  With the
construction of a fractal $N$-dimensional spacetime, I view this as
a quantum gravity effect.  As an object accelerates its fractal
geometry changes [by means of BM], thus resulting in pressure on
the system. Pressures as the source of a gravitational field were
pioneered long ago by Einstein \ct{30}:
\begin{eqnarray}
R_{\mu\nu}=-\kappa(T_{\mu\nu}-{1\over 2}g_{\mu\nu}T) \\
+{2\over a^2}\gamma_{\mu\nu}=\kappa({\sigma\over 2}-p) \\
0=-\kappa({\sigma\over 2}+ p)
\end{eqnarray}
This method is not ad hoc, gravitational pressures for atomic
gases and radiation can be given by \ct{37}:
\be
p_{gas}={\kappa\over\mu H}qT\qquad {\rm and}\qquad p_{rad}={1\over3}aT^4
\ee
which lends the general results
\begin{eqnarray}
T=\left( {\kappa\over\mu H}{3\over q}{1-\beta\over\beta}\right) ^{1/3}q^{1/3} \\
p=\left[ \left( {\kappa\over\mu H}\right) ^4{3\over a}{1-\beta\over\beta ^4}\right]^{1/3}\\q^{4/3}=c(\beta)q^{4/3}
\end{eqnarray}

Thus after a slight modification of eq.(\ref{wfe}), one can obtain
the following gravitational pressure:
\be
R^{ca}_{b}=-\kappa(T^{ca}_{b}-{1\over 2}g^{c\sigma}_{ab}T)+{2\over
a^2}\gamma^c_{ab}=\kappa({\sigma\over 2}-p) \label{rfe}
\ee
Therefore the flat field equations can be given by:
\be
\Gamma^c _{ab}(z)\ne 0,\qquad R^{ca} _{b}(z)\ne 0
\ee
thus a line elements trajectory would be given by
\be
(\omega ,z^2)=g_{ca}\omega z_{c}\omega z_{a}
\ee
therefore a metric in a complex fractal spacetime can be given by:
\be
\sum\limits_{ab=1,2}^{c=i}{\delta}\omega z_a^c \omega z_b^c\equiv 0
\ee

Since this quantum gravity effect originates from the planck
length it is very unlikely that the Yukawa interaction \ct{31}: $$
V(r)=-\int dr_1\int dr_2 {G_{\rho_1}(r_1 X r_2) \over
r_{12}}\,[r+\alpha\: exp(-r_{12}/\lambda)] $$ will take place
(unless special conditions arise).  However, if such an effect
does arise, it may yield peculiar motion for an obejects geodesic
path.

\subsection{SQM pseudo geodesic paths}\label{pgeo}
First let us begin with the two-dimensional Lagrangian
Hamiltonian, so that we have an equation of motion from the simple
action
\be
{dq_i \over dt}={\partial H \over \partial p_i};\qquad{dp_i \over dt}={\partial H \over q_i}
\ee
In canonical terms motion is given by
\be
q_i-{\partial H \over \partial p_i};\qquad p_i={\partial H \over
q_i} \ee lending a four-vector of the form \(p=\dot m x + {e \over
c} a(x)\).  In such that a Hamiltonian wave within a gravitational
field would be in motion according to the geodesic path:
\be
{\partial H \over \partial q^i}-\Gamma^{\not\alpha} _{[\not\beta
\not \gamma]}{dx^{\not\alpha} \over ds}{dx^{\not\gamma} \over
ds}\ne 0 \ee This geodesic unlike the prior for a classical
particle, will not differentiate and thus its motion need not
transverse through classical Euclidean space. Therefore it can be
seen that complex spaces could impose unseen forces which would
effect a geodesic path for a body (or wave) in motion.  A proposal
made in Ref. \ct{35}, made a like was case in the relativistic
sense so that one would have:
\be
F^{\mu}=\hat m_0{W' \over W}{dy
\over d\hat\lambda}{dx_{\mu} \over d\hat\lambda}.
\ee

Although there is no direct physical evidence of this, it is still
however an intriguing explanation.  After all the so called
``anomalous acceleration" is only experienced by small bodies, not
massive ones such as planets.  Thus a quantum interpretation of
this effect seems to fit the observed data better than any other
approach.  Alternatively Modanese has also predicted a macroscopic
quantum gravity effect \ct{32}, however it is limited to the
Podkletnov experiment \ct{9}.

\section{Discussion of theory}\label{dis}
The formulation of this theory was based on a desire for a
reformulation of GR in order to describe a singularity free
theory; in which a fractal formulation of the field equations were
derived.  The second desire for this theory was the formulation of
a quantum construction of GR, however the end result is a gravity
theory which describes quantum mechanics. Therefore the
gravitational field and matter can be considered to be molded into
the following form.  Matter exist as a pseudo point particle whos
field of movement is restricted onto a two-dimensional (complex)
frame.  This two-dimensional frame's movement is governed by BM,
and in part by the HP.  Matter, is thus in reality a fractal
vibrating string fragment which continues on into $N$-dimensions.
The fractalization of this ``cardinal string" produces virtual
particles which posses a geometry, it is this (virtual) fractal
geometry that is responsible for the gravitational field.

In light of future studies it is likely that an adequate
formulation for an alternative to GR be given in the following
forms.  One the acceptance of a fractal (even if only quasi fractal) structure of matter and space as an adequate formulation for the geometry of spacetime.  Two the acceptance of complex systems
into the equations, e.g. quaternions, octonions, C* algebras, etc.
And finally three, the acceptance of physical conditions which may
not be ``popular," but yield results that are not contradictory to
known data.  The mathematical conditions are the most intriguing
to author because there seems to be a hidden mechanism in the
mathematics.  However, my advanced mathematics skills are mediocre
at best so these avenues are still left open in this letter.

\subsection{what layith beyond the planck length?}\label{pl}
Several physical arguments against the existence of singularities
have been given by Loinger \ct{25, 26}, as well as Einstein's
classic objections.  Thus one may inquire what happens at the
planck length, i.e. what are the laws of physics?  Here I now
quote Kip Thorne, on our current understanding of singularites and
`quantum foam.'
\begin{quote}
``{\it How probable is that a black hole's singularity will give birth
to `new universes?'  We don't know.  It might well never happen,
or it might be quite common---or we might be on completly the
wrong track in believing that singularites are made of quantum
foam.}"  \flushright --Thorne (1994)
\end{quote}

This now leads a discussion to recent attempts to model gravity in terms of
$N$-dimesnional spaces (Arkani-Hamed, et al 62-69), in which the
planck length varies with the number of dimensions.  Of course the
planck length could be infinitely small in an infinite system,
clearly a challenge to the principle of planck invariance.
However, we note that with Mach's Principle (MP) the planck length
must be observed by an external mechanism to remain invariant.
Thus the planck length exist as a fractalization of BM, which
becomes an observational frame in classical real mechanics.
Furthermore, from this it may be seen that the laws of physics as
we understand them are in direct consequence of the planck length.
We may also assume the chosen string field is quantitized (i.e.
given by BM), because its mass is attributed to a complex
pseudo oscillation (vibration).  Where I now quote David Bohm (cfr. Bohm,
22):

\begin{quote}
``{\it We may conclude that all systems which oscillate are quantitized
with $E=n\hbar\nu$ whether these systems be mathematical
oscillators, sound waves, or electromagnetic waves.}"
\end{quote}
So what does physics look like beyond the planck length, remember
the (local) laws of physics are given by two complex D2-branes.
When we interpret these interactions we receive the traditional GR
effects at the macroscopic level.  However, at the planck scale
singularities don't exist such that the frame interacts via ``cardinal strings"
and not classical GR.  Thus interactions on
local branes cease, and supersymmetry takes over.  However, only
strings which are connected to a form of the D2-brane will have
observable physical manifestations, this deals with ``planckian
invariance".  In fact each dimension may have its own unique
planck length which governs its own local laws (explaining the
limitation of classical string theory to a set number of
dimensions).  Which leaves open several areas in $N$-dimensional
black hole mechanics, and planck length physics.

\section{Conclusion}\label{con}
I have shown that there is enough evidence at present to challenge
GR as the correct theory for gravitation.  I have also introduced
the study of a complex fractal spacetime system and its possible
relationship to the planck length.  The given formulation for a
canonical gravitational field resulted in contradictory
conclusions, thus ruling out a canonical approach to ``quantum
gravity."  Finally if my hypothesizes hold valid then SQM will
begin to make invalid predictions for the behavior of particles
near the planck length. Thus a fractal correction for SQM will be
needed under certain gravitational fields, which may be comparable
to BM.

\appendix
\section{the equivalence principle}\label{ep}
A new Weak Equivalence Principle (WEP) for the gravitational field
can be postulated utilizing a Complex Fractal Minkowski Spacetime
(CFMS) system (see eq.(\ref{k})).  Since there is no spatial
acceleration for the gravitational field (in respect to MP), it is
the acceleration of the pseudo time dimension in the CFMS which
produces a gravitational curvature. Therefore a material body
would have the traditional Minkowski spacetime, acting as a
Lorentz frame. Since gravitational fields extend indefinitely,
this should cause time to continually progress within an inertial
acceleration frame. This therefore means that as an object enters
a gravitational field it becomes less massive, in terms of a
Lorentz transformation.  Equivocally it can be stated that energy
is lost in curved spacetime.  A similar effect is all ready known,
known as a ``gravitational time delay," i.e. the \it Shapiro
Effect\rm .
\begin{quote}
``{\dots\it according to the general theory, the speed of a light wave
depends on the strength of the gravitational potential along its
path.}" \flushright --Shapiro (1964)
\end{quote}
This is however contradictory to SR, because it fails to describe
inertial acceleration within a gravitational field correctly.
However, the \it principle of relativity \rm is still preserved,
because the curvature of spacetime corrects for the CFMS.
Therefore the reason the \it equivalence principle \rm is
fundamental in GR is because it is the only priori condition which
satisfies the \it principle of relativity\rm .  Thus without an
equivalence principle, there would be no relativistic theory for
the gravitational field.

The logarithm gravitational time delay, may also be responsible
for the apparent ``anomalous acceleration" of spacecraft.  David
Crawford has offered a similar explanation, where the
gravitational term arises from interplanetary dust \ct{27}.

\section{Yang-Mills gravity}\label{ymg}
Here I now hit upon a topic hinted upon in section \ref{qedf};
converting fractal geometry in the terminology of QCD.  Let us now
rewrite eq.(\ref{rfe}), so that we have an equation of the form:
\be
-\kappa(R_b ^{ca}-{1\over 2}g^{c\sigma}_{ab}R^c
)={8\pi\over\sqrt{-\tilde g}}T^{c\sigma}_{ab} \ee with this
equation a Yang-Mills gravitational pressure can arise under the
following field:
\be
S_E={1\over 4g^2}\int\omega^4 zF^\sigma _{\mu\nu}F^\sigma _{\mu\nu}+{1\over\alpha _o}\int K^{ia} _b K^{ib} _a \sqrt{-\tilde g}\omega ^2 \phi
\ee
which must be given in a conformal field, i.e. $g_{ab}=p\delta_{ab}$, thus we have:
\be
S={1\over 2\alpha_0}\left[\int\omega c^2 \phi
{p^{-1}(\phi)(\partial ^2 z)^2 +\lambda ^{ab}(\partial
_{az}\partial _{bz}-p\delta_{ab})}\right] +\aleph _0\int _p \omega
^2 \phi \ee From this it is now seen that a ``cardinal string," is
in fact an $N$-dimensional world line.  Which can communacate with
other world lines, where we have a self-organization of the system
by
\be
v(\tilde s)={1\over 4\pi}\int\omega ^2 \phi\sqrt{\tilde
g}g_{ab}\epsilon^{jklm}\partial _a t_{ab}\partial_{jklm} \ee It is
these interactions which generate a spinor space, which attributes
mass to the geometry.  Therefore the Yang-Mills field is added to
the gravitational field, by means of a gravitational pressure.
Here a less restrictive form of the Strong Equivalence Principle (SEP) can be
applied:
\begin{center}
(spinor pressure)$\cdot$(energy density)\\=(strength of gravitational field)$\cdot$(gravitational pressure).
\end{center}
Thus the interaction of
two or more ``cardinal strings," produces a twistor like action,
represented by $Z^{\infty}\; at\; P\in\cal M$.  This also means
that certain gravitational anomalies may not only arise at the
planck length, and may result in experimental verification.

\bibliographystyle{plain}

\end{document}